# Ultrastrong coupling between nanoparticle plasmons and cavity photons at ambient conditions

Denis G. Baranov[1], Battulga Munkhbat[1], Elena Zhukova[2], Benjamin Rousseaux[3], Ankit Bisht[1], Adriana Canales[1], Göran Johansson[3], Tomasz Antosiewicz[1,4] and Timur Shegai[1,*]

[1]*Department of Physics, Chalmers University of Technology, 412 96, Göteborg, Sweden*

[2]*Moscow Institute of Physics and Technology, Dolgoprudny 141700, Moscow, Russia*

[3]*Department of Microtechnology and Nanoscience - MC2, Chalmers University of Technology, 412 96 Göteborg, Sweden*

[4]*Faculty of Physics, University of Warsaw, Pasteura 5, 02-093 Warsaw, Poland.*

[*]Email: timurs@chalmers.se

**Abstract:**

Ultrastrong coupling is a distinct regime of electromagnetic interaction that enables a rich variety of intriguing physical phenomena. Traditionally, this regime has been reached by coupling intersubband transitions of multiple quantum wells, superconducting artificial atoms, or two-dimensional electron gases to microcavity resonators. However, employing these platforms requires demanding experimental conditions such as cryogenic temperatures, strong magnetic fields, and high vacuum. Here, we use plasmonic nanorods array positioned at the antinode of the resonant optical Fabry-Pérot microcavity to reach the ultrastrong coupling (USC) regime at ambient conditions and without the use of magnetic fields. From optical measurements we extract the value of the interaction strength over the transition energy as high as $g/\omega \sim 0.55$, deep in the USC regime, while the nanorods array occupies only ~4% of the cavity volume. Moreover, by comparing the resonant energies of the coupled and uncoupled systems, we indirectly observe up to ~10% modification of the ground-state energy, which is a hallmark of USC. Our results suggest that plasmon-microcavity polaritons are a promising new platform for room-temperature USC realizations in the optical and infrared range.

KEYWORDS: *Ultrastrong coupling, plasmonic nanoparticle arrays, Fabry-Pérot microcavity*

Ultrastrong coupling (USC) is a regime of light-matter interaction in which the coupling strength, $g$, exceeds about ~10% of the transition energy, $\omega$ [1,2]. In this regime, the standard quantum optical approximations, such as the commonly made rotating wave approximation (RWA), fail. Thus the fast rotating terms, as well as the quadratic $A^2$ term must be taken into account in order to correctly describe the system's behavior [3–5]. Remarkably, not only quantum two-level systems, but also classical harmonic oscillators in the regime of ultrastrong coupling require description using the full Hamiltonians [6]. The intriguing result of USC is that the global ground state of the system gains a photonic component, that is, the ground state contains a finite number of virtual photon excitations [7,8]. This in turn may lead to highly unusual phenomena, such as dynamical Casimir effect [9–11] and single-photon frequency conversion [12]. The $A^2$ term may furthermore reduce the dipole-field interaction due to effective screening of the dipoles from the field [13].

Although the USC domain of light-matter interaction is of significant fundamental interest, it remains largely unexplored experimentally due to technical challenges of its realization. Indeed, so far the record-high realizations (where $g/\omega > 1$) have been based on Landau polaritons [14] and superconducting circuits [15], which require cryogenic temperatures and high magnetic fields. This specific interaction regime for which $g/\omega > 1$ is called "deep" strong coupling. However, replicating such results under ambient conditions remains a challenge. Room temperature realizations using collective coupling of organic molecules with microcavities have reached $g/\omega$ of "only" ~0.3 [16,17], with the recent implementation based on intersubband transitions of heavily doped quantum wells showing $g/\omega$~0.45 [18]. Plasmonic lattices [19,20] as well as single plasmonic nanorods [21] have been shown to couple strongly with microcavity modes previously, however, the reported interaction strengths have not reached the level of the USC regime.

Here, we use our recently developed strategy based on plasmon-microcavity polaritons [22] to achieve considerably higher coupling strengths, well into the USC regime. Our system is scalable, engineerable and highly controllable, thus offering a unique platform for realization of USC regime at ambient conditions.

**Ultrastrong coupling in plasmon-microcavity systems**

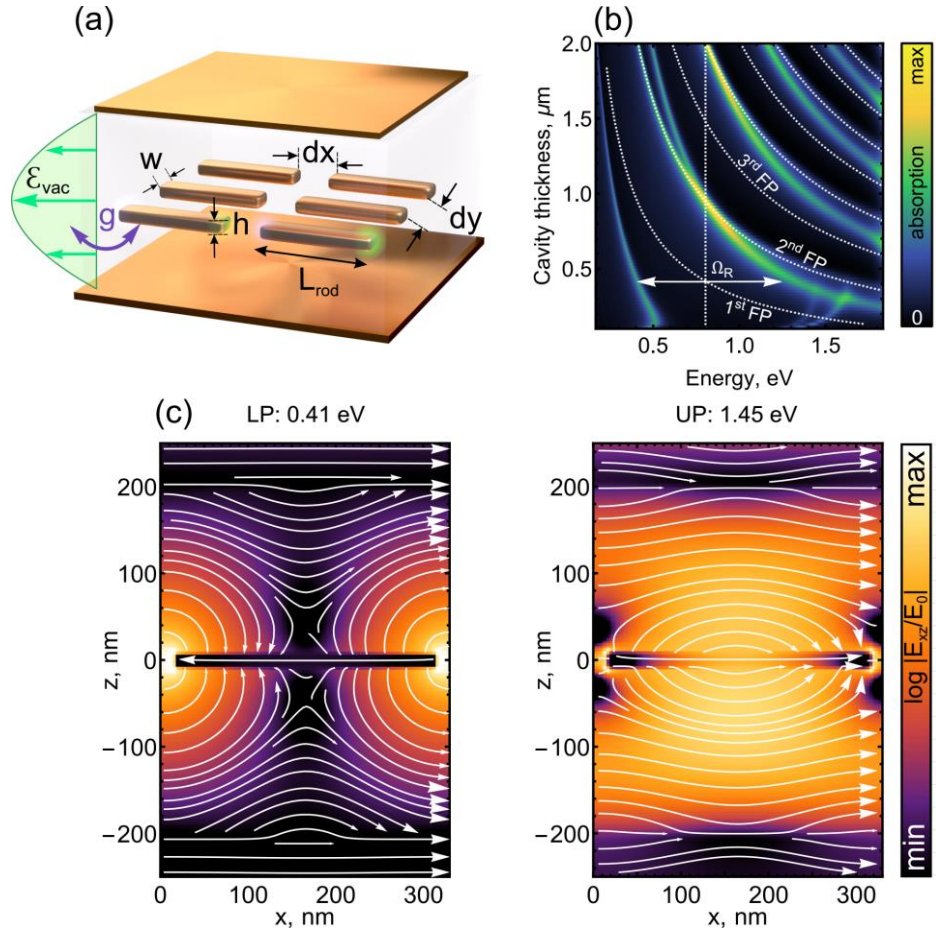

**Fig. 1.** (a) Artistic illustration of the system: an array of plasmonic nanorods positioned in the middle of a Fabry-Pérot cavity formed by two gold mirrors. The cavity interior is filled with $SiO_2$. The array couples to the FP cavity mode, exchanging energy at a rate $g$. (b) False-color normal-incidence absorption spectra as a function of cavity thickness with an array of 300 nm long plasmonic nanorod (width $w = 50$ nm, height $h = 20$ nm) positioned in the middle of $SiO_2$-filled Fabry-Pérot cavity. The vertical dashed line indicates the nanorod plasmon resonance outside of the cavity. The curved lines indicate resonances of the empty FP cavity, whose even modes are not modified by the coupling. $\Omega_R$ denotes plasmon-cavity mode splitting at zero detuning (c) The electric field intensity (in the log scale) and the electric field lines in the vertical plane across the middle of the nanorod induced by a normally incident plane wave (polarized in the figure plane) for the coupled system of 400 nm thick cavity and 300 nm long nanorods calculated for the lower and upper polaritons.

The system under study is illustrated in Fig. 1a. It consists of a sub-diffractive periodic array of Au nanorods placed in the antinode of the fundamental Fabry-Pérot (FP) microcavity mode formed by two gold (Au) mirrors and filled by a $SiO_2$ spacer. The nanorods array couples to the vacuum field of the FP microcavity, thus producing plasmon-cavity polaritons manifested as distinct resonant spectral features emerging in transmission, reflection and absorption spectra of the coupled system.

To provide initial insight into the behavior of the coupled system, we perform numerical finite-difference time-domain (FDTD) simulations (see Methods). Fig. 1b shows a map of absorption spectra of coupled FP-nanorod systems at normal incidence with the electric field parallel to the nanowires as a function of the cavity thickness for nanorod lengths $L = 300$ nm and $dx = dy = 30$ nm spacing. The bare FP cavity resonances are shown as dashed curves. The vertical dashed line marks the bare plasmon resonance of the array. In the coupled system, we observe an emergence of new eigenmodes – with the even FP modes being practically unperturbed, while the odd FP modes being significantly hybridized with plasmon modes.

The 1$^{st}$ order FP mode of an empty cavity intersects the bare nanorod array plasmon resonance around 400 nm cavity thickness resulting in a distinct anticrossing, Fig. 1b. The lower polariton (LP) transitions from a plasmon-dominated mode (for a thin cavity) to an FP-dominated mode at large detuning (for a thick cavity). However, the upper polariton (UP) upon acquiring a plasmon-like character at large detuning, crosses the 2$^{nd}$ order FP mode and approaches the spectral position of the 3$^{rd}$ FP cavity mode, which is strongly pushed to the blue due to hybridization with the plasmon. Such qualitative blue shift behavior is observed for all odd modes. In contrast, the even modes do not significantly couple to the array due to symmetry. These observations suggest that a multimode character of the FP microcavity is important for a detailed interpretation of our results.

The spatial distribution of the electric field induced by a normally incident plane wave inside the plasmon-cavity system calculated at the resonant energies for a 400 nm thick cavity clearly displays the opposite symmetries of the two resonances, Fig. 1c. While the lower energy mode shows an anti-symmetric combination of cavity and plasmon fields, featuring two saddle points above and below the nanorod, the upper energy mode is a symmetric combination. Such behavior highlights the polaritonic nature of the two resonances of the hybrid system. For a 400 nm thick cavity, corresponding to near-resonant coupling ($\omega_{cav} = \omega_{pl} \sim 0.8$ eV), the Rabi splitting, $\Omega_R$, estimated as the energy difference between the two absorption peaks reaches $\sim 1$ eV. Thus, assuming that $\Omega_R = 2g$ at resonance, we estimate the normalized coupling strength of $g/\omega_{pl} > 0.5$, which clearly indicates the ultrastrong coupling regime in the system. In what follows, we perform a more rigorous estimation of the $g/\omega_{pl}$ values in our systems based on a full Hopfield Hamiltonian.

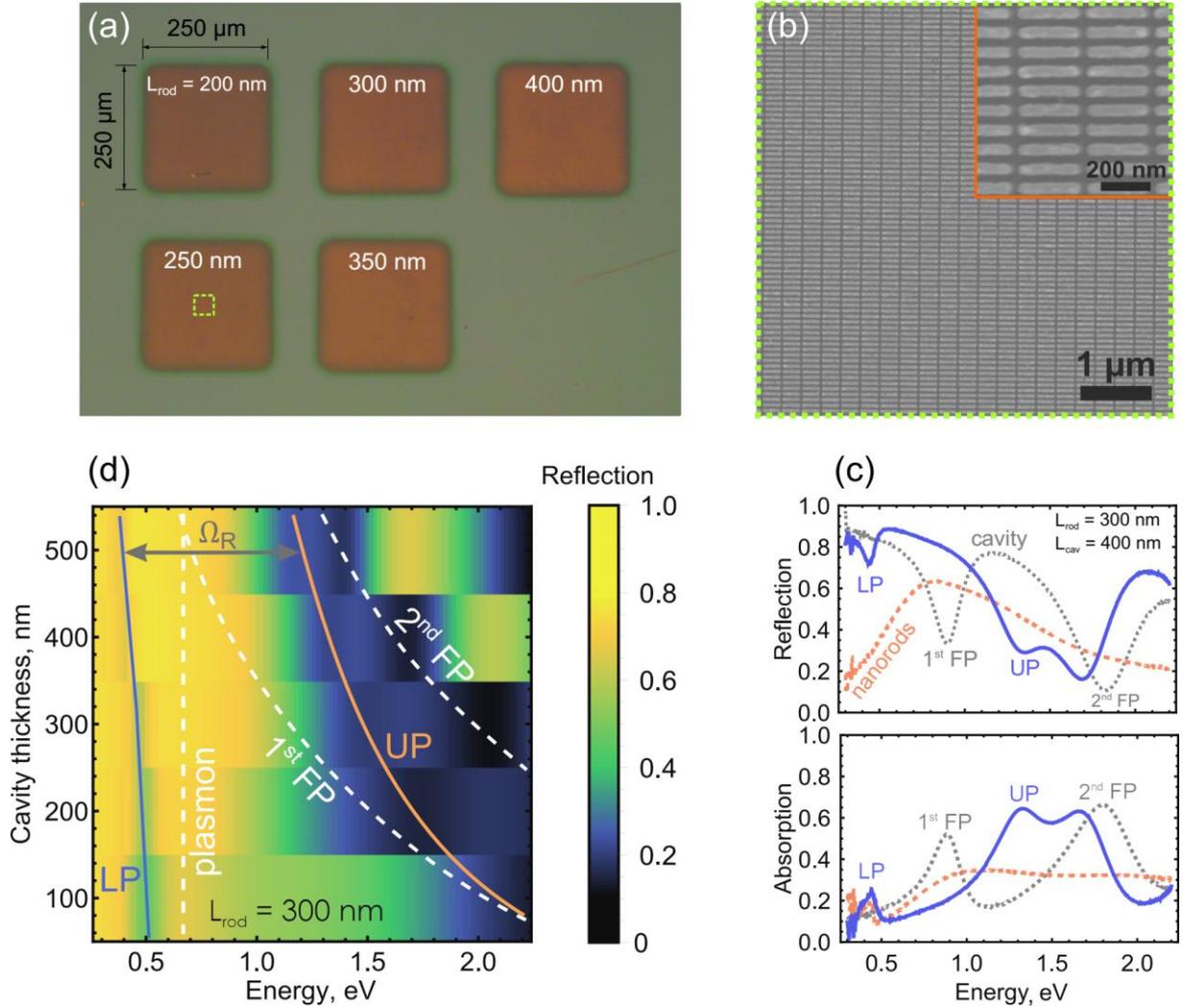

**Fig. 2.** (a) Bright-field optical microscope images of gold nanorod arrays positioned in the middle of a SiO$_2$-filled FP cavity (without the top mirror) fabricated by electron beam lithography. Individual nanorods have variable length 200 to 400 nm ($w = 50$ nm, $h = 20$ nm). The side-to-side distance between the nanorods is 30 nm. The arrays are 250×250 μm$^2$. (b) SEM image of the $L_{rod} = 250$ nm nanorods array. The inset shows a magnified view of the nanorod array. (c) Measured reflection (a) and absorption (b) spectra of an empty $L_{cav} = 400$ nm cavity, bare $L_{rod} = 300$ nm long plasmonic nanorods, and those of the coupled system with the electric field polarization parallel to the major rod axis. (d) Measured dispersion of the reflection spectra of the coupled plasmon-cavity system with $L_{rod} = 300$ nm plasmonic nanorods as a function of the cavity thickness revealing an anti-crossing between the two polaritonic modes.

Samples of coupled plasmon-microcavity systems were fabricated by a combination of electron beam evaporation (Au mirrors), plasma-enhanced chemical vapor deposition (dielectric spacers), and electron beam lithography (nanorod arrays) (see Methods). Fig. 2a shows a bright-field optical image of the nanorod arrays with lengths ranging from 200 nm to 400 nm. The cavity thicknesses ranged from 100 to 500 nm. The nanorods have fixed widths of $w = 50$ nm and heights of $h = 20$ nm, which accounts for filling only ~4% of the resonant cavity interior. An SEM image of $L_{rod} = 250$ nm gold nanorods array is shown in Fig. 2b (see Methods). Both figures clearly show high-density plasmonic arrays with an interparticle distance as small as 30 nm, corresponding to the area filling factor of 60%. More examples are shown in Fig. S6.

Next, we proceed to optical measurements of the fabricated plasmon-cavity systems using the Fourier transform infrared (FTIR) spectroscopy (see Methods). Figs. 2c show normal incidence reflection and absorption spectra of an empty 400 nm thick cavity, 300 nm long nanorods array, and those of the coupled system (Figs. S7-8 show uncoupled data). The uncoupled cavity and array resonances overlap spectrally and, when coupled, unambiguously confirm the realization of a giant Rabi splitting in the spectra of the coupled plasmon-cavity systems.

Dispersion of measured normal-incidence reflection spectra from coupled systems with 300 nm long nanorods and varying cavity thickness displays a clear anticrossing between the 1$^{st}$ order FP mode and the plasmon mode of the array, Fig. 2d (see also Fig. S9). The spectra also reveal the 2$^{nd}$ order FP mode (third dip from the left), which does not interact with the nanorods due to the electric field node in the center of the cavity. Based on these spectra, the vacuum Rabi splitting taken as the energy difference between the two reflection dips at zero detuning ($\omega_{cav} = \omega_{pl}$, 500 nm thick cavity), reaches ~0.8 eV at the resonant energy of ~0.7 eV, Fig. 2d. Thus, the Rabi splitting in our samples exceeds both the bare cavity and bare plasmon resonance frequencies, indicating that the hybrid plasmon-cavity system is deep into the USC regime.

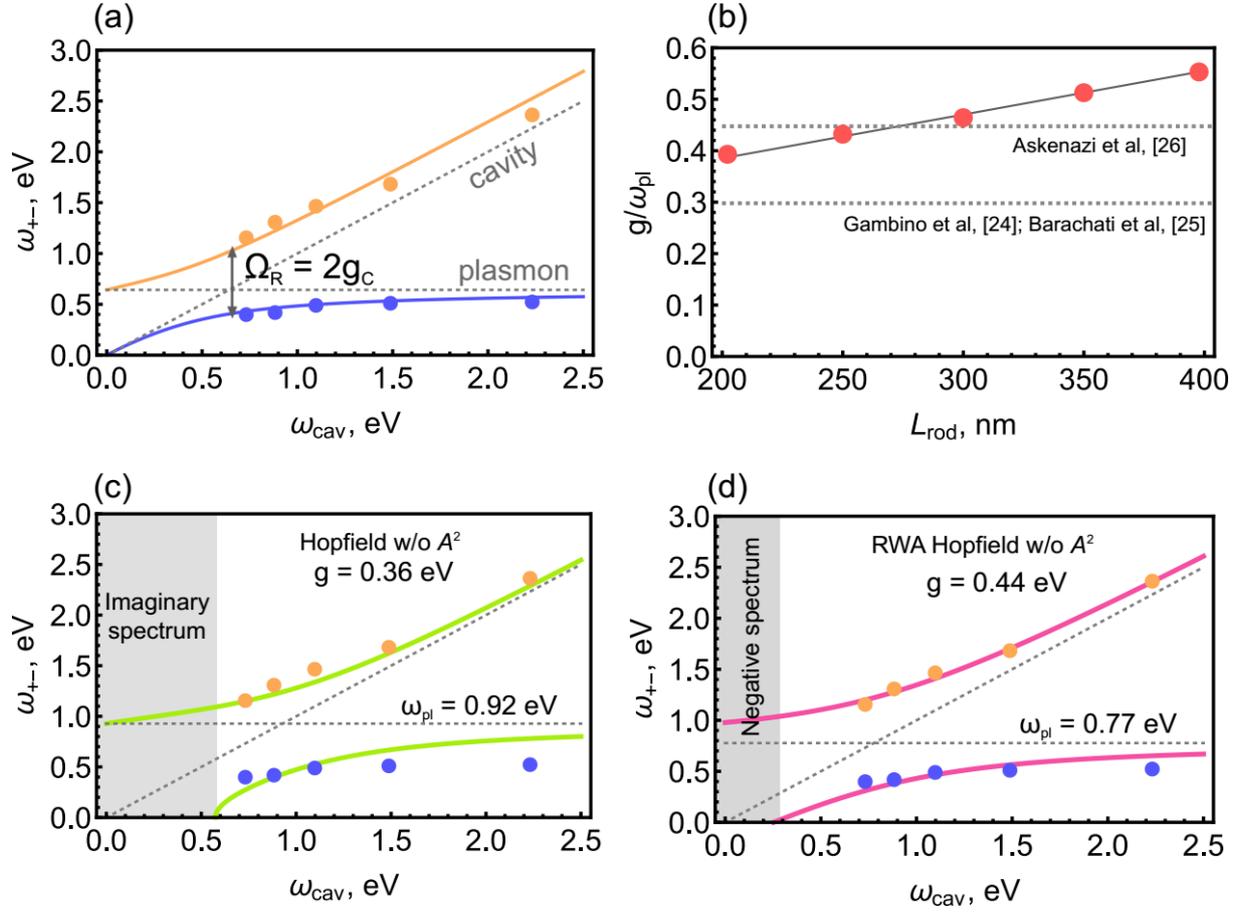

**Fig. 3**. (a) Fitting of the measured polaritonic dispersion of the coupled plasmon-cavity system ($L_{rod} = 300$ nm) with Hopfield Hamiltonian transition energies. Dots show resonant energies of the coupled system extracted as experimental reflection dips, lines are Hopfield polaritons dispersion, gray dashed lines are the bare cavity and bare plasmon energies. (b) Normalized coupling strength at zero detuning versus nanorod length obtained in this work compared to previous state-of-the-art results. (c) Fitting of the measured polaritonic dispersion of the coupled plasmon-cavity system using the single-mode Hopfield Hamiltonian without the $A^2$ term results in unphysical imaginary energies of the lower polariton. (d) Fitting the same data using the Hopfield Hamiltonian under rotating wave approximation (RWA) and without the $A^2$ term results in negative energies of the lower polariton.

**Analysis of the ultrastrong coupling using Hopfield Hamiltonian**

We now turn to a more thorough analysis of the experimental data. Since a rough estimation already reveals that the Rabi splitting in our system is comparable to the transition energy of uncoupled oscillators, the usual Jaynes-Cummings or Rabi-type coupled Hamiltonians are invalid, and a more general Hamiltonian must be used. We thus employ the full Hopfield Hamiltonian formalism, which includes both the fast-rotating and the quadratic $A^2$ term [1]. We focus on the two lowest plasmon-cavity modes, hence we consider the coupling between two oscillators: the 1$^{st}$ order normal incidence FP mode of the cavity and the collective long-axis plasmon mode of the array. The Hamiltonian thus reads:

$$\hat{H} = \hbar\omega_{cav}\left(\frac{1}{2} + \hat{a}^\dagger\hat{a}\right) + \hbar\omega_{pl}\left(\frac{1}{2} + \hat{b}^\dagger\hat{b}\right) + \hat{H}_{int}, \qquad (1)$$

where $\hat{a}$ and $\hat{b}$ are the microcavity and collective plasmon annihilation operators respectively, and $\hat{H}_{int}$ is the interaction Hamiltonian. If we considered individual nanoparticle plasmons interacting with each other instead of the collective array mode, the Hamiltonian would also yield additional eigenstates weakly interacting with light [8]. As long as we work away from the Rayleigh modes of the array, which is ensured by sub-diffraction periodicity, all the plasmon-plasmon interaction effects can be absorbed into the single collective plasmon frequency $\omega_{pl}$ [23]. We assume that this collective plasmon frequency is the same in free space and inside the FP cavity.

The Hamiltonian can be written differently depending on the gauge in which the electromagnetic field is treated. The two options that are often used are the Coulomb gauge and its dipole representation. The latter can be obtained from the Coulomb representation by performing the Power-Zienau-Woolley (PZW) transformation [24]. When a cavity couples to a two-level system, the two representations are not invariant because of the two-level approximation [25,26]. However, since we are considering coupling of two harmonic oscillators, the two representations provide identical spectra [6,8]. We therefore will use the Coulomb representation, in which the single-mode interaction Hamiltonian can be written as [13,27]:

$$\hat{H}_{int} = \hbar g_C(\hat{a}^\dagger + \hat{a})(\hat{b}^\dagger + \hat{b}) + \frac{\hbar g_C^2}{\omega_{pl}}(\hat{a}^\dagger + \hat{a})^2, \qquad (2)$$

where $\hbar g_C = \mu_{pl}\sqrt{a^2\rho}\,\varepsilon_{vac}\frac{\omega_{pl}}{\omega_{cav}}$ is the coupling strength with $\mu_{pl}$ being the transition dipole moment of the plasmonic nanorod, $\rho$ the plasmonic nanoparticles density per unit area $a^2$ and $\varepsilon_{vac} = \sqrt{\frac{\hbar\omega_{cav}}{2\varepsilon\varepsilon_0 a^2 L_{eff}}}$ the vacuum electric field of the cavity with $L_{eff}$ being the effective cavity mode transverse thickness [8]. The first term in Eq. (2) is the usual Rabi-type interaction including both slow and fast-rotating terms. The second term is the so-called $A^2$ term, which arises from expansion of the minimal coupling Hamiltonian $\left(\mathbf{p} - \frac{e}{c}\mathbf{A}\right)^2$ and "protects" the coupled system from the superradiant phase transition [4,5], as well as stabilizes the spectrum against the square-root singularity [28]. Neglecting the $A^2$ term leads to the breakdown of the entire energy spectrum at $g_C = \omega_{cav}/2$, Fig. S12. Neglecting additionally the fast-rotating terms yields a Jaynes-Cummings-like spectrum with the superradiant phase transition at large coupling strength, Fig. S12.

In a classical optical experiment, such as elastic scattering, reflection, or absorption, one cannot access the ground-state energy directly. However, spectral positions of the resonant features in reflection or absorption spectra reflect approximately the transition energies between the ground and first excited states of the system $\hbar\omega_\pm = E_{\pm 1} - E_0$, Fig. S12. Therefore, to model the system with the Hopfield Hamiltonian framework, we fit the measured dispersions of reflection dips with calculated transition energies $\hbar\omega_\pm$ of the Hopfield Hamiltonian [27]. The spectrum of transition energies of Hamiltonian (1, 2) can be obtained as solutions of the Hopfield problem (see Methods).

The resulting Hamiltonian fit of a coupled system's resonant transitions as a function of the bare cavity energy is presented in Fig. 3a for $L_{rod} = 300$ nm nanorod arrays. For each cavity thickness, the bare cavity energy was obtained from the reflection (Fig. S7). By assuming that the effective cavity thickness scales as $L_{eff} = \frac{\lambda_{cav}}{4n}$ with $n$ being refractive index of the cavity medium, we arrive at the coupling strength in the Coulomb representation $g_C = \omega_{pl}\mu_{pl}\sqrt{\frac{\hbar\rho}{\pi\varepsilon_0 nc}}$, which is independent of the cavity thickness and energy. Hence, we fit the polaritonic dispersion by freely varying plasmon resonance frequency $\omega_{pl}$ and the coupling strength $g_C$.

For the $L_{rod} = 300$ nm nanorod arrays, the fitting yields the plasmon frequency of 640 meV and the coupling strength of 300 meV (see Fig. S13 and Table SI). For all 5 nanorod lengths, we consistently obtain normalized coupling strength values $g_C/\omega_{pl}$ in the range from 0.4 to 0.56, Fig. 3b, which unambiguously indicates the USC regime of interaction between the nanorods and the cavity modes and sets the record for room-temperature implementations of ultrastrongly coupled systems [1]. Furthermore, we notice that the normalized coupling strength $g_C/\omega_{pl} = \mu_{pl}\sqrt{\frac{\hbar\rho}{\pi\varepsilon_0 nc}}$ is a function of the effective dipole moment and the particles density only. Therefore, if the product $\mu_{pl}\sqrt{\rho}$ grows with increasing nanorod length, we may expect even higher values of $g_C/\omega_{pl}$ for longer rods resonating at lower energies.

We also compare the resulting fits with those obtained by applying the multimode Hopfield Hamiltonian accounting for all the normal-incidence modes of the FP cavity, which can be solved analytically [13] (see Fig. S14). The comparison shows no significant deviations between single-mode and multimode approaches at the range of parameters used.

We further illustrate the importance of keeping the quadratic term by analyzing the data with an *a priori* incorrect Hamiltonians. Fitting the experimental data with eigenvalues of Hopfield Hamiltonian without the $A^2$ term yields a spectrum with imaginary energies and slightly overestimated coupling strength of ~0.36 eV, Fig. 3c. This imaginary spectrum is a fundamental property of the coupled oscillators Hamiltonian without any kind of quadratic stabilizing term [28,29]. Fitting the data with no $A^2$ Hopfield Hamiltonian under RWA (without fast-rotating terms), although appears to yield a better fit, yields a region with negative LP energy and a largely overestimated coupling strength of ~0.44 eV, Fig. 3d. Similar behavior was observed when we fitted other data sets with $L_{rod} = 200$ and 400 nm with incomplete Hamiltonians, Fig. S17.

**Ground-state energy and photonic occupancy**

Having performed the fitting of the experimental data, we can analyze how the ground state of the system $|G\rangle$ is modified by the ultrastrong coupling. In the uncoupled case, the global ground state is a direct product of the zero-photon and zero-plasmon states $|G\rangle = |0_{cav}\rangle \otimes |0_{pl}\rangle$, and the energy of this state is $E_G = \langle 0|H_{cav} + H_{pl}|0\rangle = \frac{\hbar}{2}(\omega_{cav} + \omega_{pl})$, correspondingly. The USC modifies the global ground state $|\tilde{G}\rangle$ by admixing the states with different number of excitations, i.e. the global ground state with the higher excited states [7], thus modifying the ground-state energy. Since after diagonalization, the coupled system comprises two new harmonic oscillators, its ground-state energy is $\tilde{E}_G = \frac{\hbar}{2}(\omega_+ + \omega_-)$.

The ground-state energy change, $\delta E_G = \tilde{E}_G - E_G$, at zero cavity-plasmon detuning can be estimated as $\delta E_G \approx \frac{g_C^2}{2\omega_{cav}}$ (see Methods), which, for $g_C/\omega_{pl} \approx 0.5$ as in our case, yields $\delta E_G \approx g_C/4 \approx 75$ meV accounting for about 12% of the unperturbed ground-state energy $E_G$. Thus, the absolute ground-state energy change in our system is several times greater than $k_B T$ at room temperature. We stress that such ground-state energy modification is significant and thus may show up in practical USC-related effects even at room temperature.

The normalized ground-state energy variation $\frac{\delta E_G}{E_G} = \frac{\tilde{E}_G - E_G}{E_G} = \frac{\omega_+ + \omega_-}{\omega_{cav} + \omega_{pl}} - 1$ calculated using the obtained coupling strengths and analytical expressions for polariton energies $\omega_\pm$, Fig. 4a, predicts up to ~10% modification of the ground-state energy for normal incidence FP mode upon coupling with the plasmonic array (see also Fig. S18).

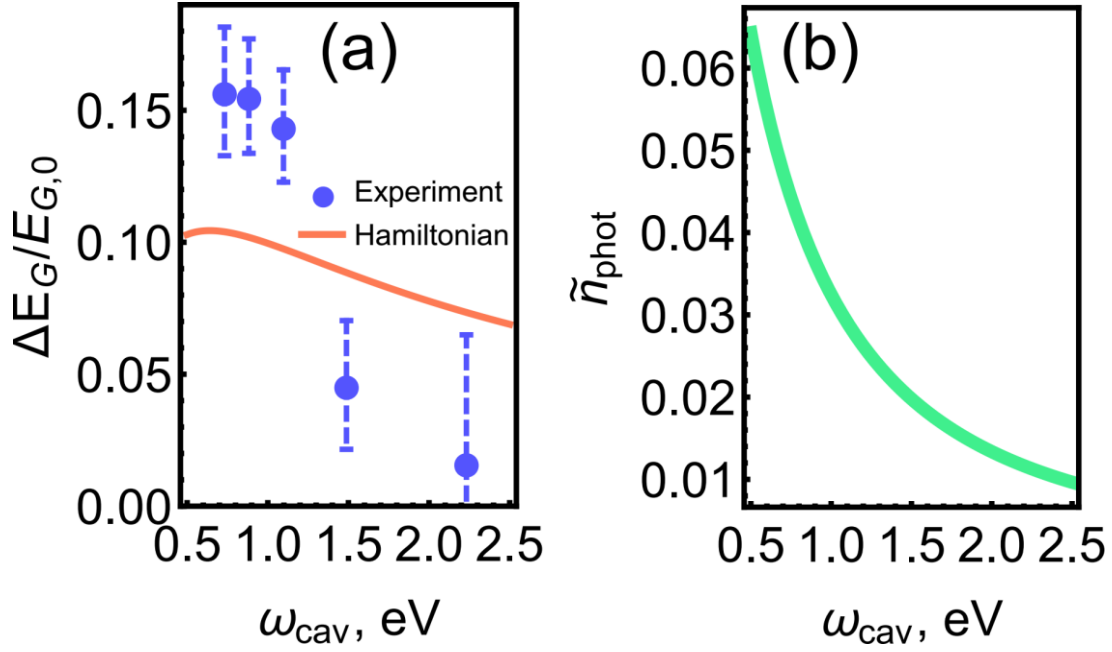

**Fig. 4**. Modification of the vacuum state by the ultrastrong coupling. (a) The normalized vacuum energy variation in the coupled plasmon-cavity system as a function of the bare cavity energy for normal incidence eigenmodes calculated with the coupling strength obtained from fitting of the $L_{rod} = 300$ nm system, as well as the vacuum energy variation calculated directly from the measured polariton energies. (b) The photonic occupancy of the modified ground state in the coupled system as a function of the bare cavity energy calculated with the coupling strength obtained from fitting of the $L_{rod} = 300$ nm system.

These theoretical values follow the experimental trend (circles in Fig. 4a), which was obtained using the measured cavity and polariton energies with only the bare plasmon frequency $\omega_{pl}$ adopted from the fitting. The theory predicts a relatively slow dependence of the normalized ground-state energy change on the detuning, whereas the experiment is more sensitive to that. This can be explained by the error in determination of polaritons energies. Despite the non-ideal agreement, however, we stress that both theoretical predictions and experimental reflectivity data signal the ground-state energy modification of the order of 10% in our plasmon-microcavity systems. Such a modification is a clear hallmark of ultrastrong coupling, since in the conventional strong coupling picture, where $g \ll \omega$, the additive coupled and uncoupled energies are exactly the same, i.e. $\omega_+ + \omega_- = \omega_{cav} + \omega_{pl}$, as can be seen from the Jaynes-Cummings model.

Lastly, we study the photonic occupancy $\tilde{n}_{phot} = \langle \tilde{G} | \hat{a}^\dagger \hat{a} | \tilde{G} \rangle$ of the modified ground state $|\tilde{G}\rangle$ (the plasmon occupancy of the ground state $\langle \tilde{G} | \hat{b}^\dagger \hat{b} | \tilde{G} \rangle$ equals the photonic one [7]). In the USC regime, the ground state of the system acquires a non-zero photonic component due to the aforementioned admixing of states with different excitation numbers [7]. The photonic occupancy for the $L_{rod} = 300$ nm coupled systems, Fig. 4b, suggests that the ground state of the ultrastrongly coupled system contains up to 0.06 bare cavity photons for cavities resonant with the nanorod array ($\omega_{pl} = \omega_{cav} \sim 0.5$ eV). This is smaller than 0.37 photons estimated for Landau polaritons in the THz range [14], but it is still a feasible number for converting to real photons by fast modulation of the coupling strength. The photonic occupancies calculated for other plasmonic nanorods predict almost identical values, Fig. S19.

**Discussion**

Above we have presented the ground-state modification taking into consideration only the normal incidence ($k_\parallel = 0$) mode of the cavity, whereas in reality all cavity modes having various in-plane momenta $k_\parallel$ as well as TM and TE polarizations will couple to the nanorod array. Due to the periodicity of the system, modes with different $k_\parallel$ do not interact and can be treated with independent Hamiltonians. The full vacuum energy per unit area of the cavity therefore can be calculated by integrating the vacuum energy over the entire $k$-space of the system. However, such an integration will diverge due to the asymptotic growth of the FP modes energy at large $k_\parallel$. A regularization scheme will likely be needed to obtain a finite value similarly to the well-known result of Casimir [30]; these calculations will be considered elsewhere.

Our plasmon-microcavity system offers a number of interesting perspectives. First, we have studied coupled systems with only one layer of plasmonic nanoparticles that occupies only ~4% of the cavity interior. However, one can readily scale up the process and place several plasmonic layers in the center of the cavity close to the electric field anti-node. For example, placing 4 identical layers, assuming they all interact with the maximal electric field, will double the coupling strength and enable deep ultrastrong coupling with $g_C/\omega_{pl} > 1$. Second, we note that the extracted value of $g_C/\omega_{pl}$ monotonically increases with the nanorod length in the range of studied parameters, Fig. 3b. It is therefore interesting whether the normalized coupling strength can be further boosted by increasing the nanorods length, and at which rod length the maximal $g_C/\omega_{pl}$ ratio can be expected? As we showed above, $g_C/\omega_{pl}$ in our system scales as $\mu_{pl}\sqrt{\rho}$, which likely has an optimum. Furthermore, by controlling the nanoparticles density, which is straightforward using the electron beam lithography, our system allows creating a vacuum energy gradient in the lateral direction. Lastly, the nanoparticles can be made chiral [31], opening the opportunities to create chiral vacuum states with various vacuum energies depending on the handedness of the chiral meta-atom.

To conclude, we have demonstrated a room-temperature ultrastrong coupling between two optical harmonic oscillators: a Fabry-Pérot microcavity and an array of plasmonic nanorods. The coupling strength reaches more than half of the cavity transition energy, thus unambiguously indicating the USC regime and setting the record-high value for room-temperature implementations of $g_C/\omega_{pl} > 0.55$. Analysis of the experimental data by a Hopfield Hamiltonian reveals significant deviation of the coupled system's eigenenergies from those predicted by the naïve coupled oscillators model. Remarkably, the naïve models fail to describe our system despite its obvious classical nature – both system's components, plasmonic arrays and Au mirrors contain millions of electrons and thus can be treated as classical harmonic oscillators. Furthermore, we indirectly observe a modification of the ground-state energy (up to 10%) and associated with that finite photonic occupancy induced by USC. Our findings thus introduce a plasmon-microcavity system as a new and promising platform for studies of USC and related phenomena in the optical and infrared range at ambient conditions.

**Methods**

**Samples fabrication.** All samples were prepared on thin microscope glass (170 µm) coverslips. The glass coverslips were cleaned in acetone and isopropanol at 60°C in ultrasonicator, dried with $N_2$ blow, followed by oxygen plasma cleaning. Subsequently, 10 nm of gold (Au) mirror was prepared by e-beam evaporator with adhesion layer of chromium (2 nm) to form a bottom mirror. Then, various thicknesses of $SiO_2$ layer for half-cavities were deposited by plasma-enhanced chemical vapor deposition (PECVD at 300°C) on top of a freshly-prepared bottom gold mirror.

To fabricate a coupled system, lattice arrays of gold nanorods with various sizes and densities were fabricated on top of the half-cavities using a standard e-beam lithography. Then, the top-half $SiO_2$ layers with the same thicknesses as the bottom $SiO_2$ half cavities were deposited using PECVD. Finally, the coupled samples were completed by a deposition of 10 nm gold film as a top mirror for Fabry-Pérot cavity. Bare nanorod samples were prepared directly on top of glass substrates as a reference sample. To perform further scanning electron microscopy (SEM) characterization, the samples were coated by a thin layer of conductive polymer (E-spacer). Morphology of the samples was characterized using a Zeiss (Germany) scanning electron microscope (SEM ULTRA 55 FEG).

**Optical measurements.** Infrared optical measurements were performed with a Bruker Hyperion 2000 IR microscope (Schwarzschild-objective with 15× magnification, NA = 0.4) coupled to a Fourier-transform Bruker Vertex 80v spectrometer with a liquid-nitrogen-cooled mercury cadmium telluride detector. Reflection and transmission spectra were collected at normal incidence from a sample area of about 80×80 µm$^2$ with 2 cm$^{-1}$ resolution. All spectra were obtained with $CaF_2$ IR polarizer in two principle orientations with the electric field polarization parallel and perpendicular to the nanorods long axis. A plane gold mirror was used as a reference in the reflection configuration experiment. Broad band absorption spectra were calculated from the measured reflection and transmission spectra. Reflection spectra in visible spectrum range were collected at normal incidence using a 20× magnification objective (Nikon, NA = 0.45), directed to a fiber-coupled spectrometer and normalized with reflection from a standard dielectric-coated silver mirror.

**FDTD simulations.** Finite-difference time-domain (FDTD) simulations of the electromagnetic response of the coupled plasmon-cavity system were performed using commercial software (FDTD Solutions, Lumerical, Inc., Canada). Transmission and absorption spectra, as well as electromagnetic field distributions were obtained with the use of a linearly polarized normally incident plane wave source and periodic boundary conditions with symmetries. The plane wave was polarized either along the nanorods or perpendicular to them. The permittivity of gold was approximated by interpolating the experimental data from Palik in the range 600-8000 nm. The simulation volume was discretized into a $\Delta r = 4$ nm mesh with further refinement of 2 nm around the metal structures (nanorod and both mirrors).

**Hopfield Hamiltonian diagonalization.** Spectrum of transition energies of Hamiltonian (1) with the interaction part (2) can be obtained as solutions of the following eigenproblem [27]:

$$[\hat{H}, \hat{P}] = \hbar\omega_\pm \hat{P}, \tag{3}$$

where $\hat{P} = \alpha\hat{a} + \beta\hat{b} + \gamma\hat{a}^\dagger + \delta\hat{b}^\dagger$ is the polariton operator. Rewriting the eigenproblem in the basis of $\hat{a}$, $\hat{b}$, $\hat{a}^\dagger$, and $\hat{b}^\dagger$, solutions can be found as eigenvalues of the Hopfield matrix:

$$\widehat{M} = \begin{pmatrix} \omega_{cav} + 2\frac{g_C^2}{\omega_{pl}} & -2\frac{g_C^2}{\omega_{pl}} & ig_C & -ig_C \\ 2\frac{g_C^2}{\omega_{pl}} & -\omega_{cav} - 2\frac{g_C^2}{\omega_{pl}} & -ig_C & ig_C \\ -ig_C & -ig_C & \omega_{pl} & 0 \\ -ig_C & -ig_C & 0 & -\omega_{pl} \end{pmatrix} \tag{4}$$

Two eigenvalues $\omega_\pm$ of the above matrix are given by the positive solutions of the bi-quadratic equation:

$$(\omega_\pm^2 - \omega_{cav}^2)(\omega_\pm^2 - \omega_{pl}^2) - \frac{4g_C^2\omega_\pm^2\omega_{cav}}{\omega_{pl}} = 0 \tag{5}$$

Thanks to the harmonicity of the coupled system, its entire energy ladder can be restored by collecting all possible values $\omega_{n,m} = \omega_0 + n\omega_+ + m\omega_-$ where $n$ and $m$ are non-negative integers and $\omega_0 = \frac{\omega_+ + \omega_-}{2}$.

Since the ground-state energy of a harmonic oscillator (or a set thereof) is half the transition energy (sum of those), its modification can be calculated as $\delta E_G = \frac{\hbar}{2}(\omega_+ + \omega_- - \omega_{cav} - \omega_{pl})$. By expanding the solution of Eq. (5) into a Taylor series near $g_C = 0$, the ground-state energy modification can be approximated by:

$$\delta E_G = \frac{\omega_{cav}}{\sqrt{2}\omega_{pl}} \frac{\sqrt{\omega_{cav}^2 + \omega_{pl}^2 + |\omega_{cav}^2 - \omega_{pl}^2|} - \sqrt{\omega_{cav}^2 + \omega_{pl}^2 - |\omega_{cav}^2 - \omega_{pl}^2|}}{|\omega_{cav}^2 - \omega_{pl}^2|} g_C^2 + O(g_C^4), \tag{6}$$

which at zero detuning ($\omega_{pl} = \omega_{cav}$) yields $\delta E_G = \frac{g_C^2}{2\omega_{cav}} + O(g_C^4)$.